\documentclass[aps,prb,twocolumn,nobalancelastpage,nofootinbib]{revtex4-1}

\usepackage{amsmath}    
\usepackage{amssymb}
\usepackage{graphicx}   
\usepackage{verbatim}   
\usepackage{color}      
\usepackage[dvipdfmx]{pict2e}
\usepackage{bm}
\usepackage{braket}
\usepackage{hyperref}   
\begin{document}

\title{Current-induced orbital magnetization in systems without inversion symmetry}
\author{Daisuke Hara$^1$}
\author{M. S. Bahramy$^2$}
\altaffiliation{Present address: Department of Physics and Astronomy, The University of Manchester, Oxford Road, Manchester M13 9PL, United Kingdom}
\author{Shuichi Murakami$^{1,3}$}
\affiliation{$^1$Department of Physics, Tokyo Institute of Technology, 2-12-1 Ookayama, Meguro-ku, Tokyo 152-8551, Japan}
\affiliation{$^2$Department of Applied Physics, The University of Tokyo, Tokyo 113-8656, Japan }

\affiliation{$^3$TIES, Tokyo Institute of Technology, 2-12-1 Ookayama, Meguro-ku, Tokyo 152-8551, Japan}
\date{\today}

\begin{abstract}
In systems with time-reversal symmetry, the orbital magnetization is zero in equilibrium.
Recently, it has been proposed that the orbital magnetization can be induced by an electric current in a helical crystal structure in the same manner as that in a classical solenoid.
In this paper, we extend this theory and study the current-induced orbital magnetization in a broader class of systems without inversion symmetry.
First, we consider polar metals which have no inversion symmetry.
We find that the current-induced orbital magnetization appears in a direction perpendicular to the electric current even without spin-orbit coupling.
Using the perturbation method, we physically clarify how the current-induced orbital magnetization appears in polar metals.
As an example, we calculate the current-induced orbital magnetization in SnP, and find that it might be sufficiently large for measurement.
Next, we consider a two-dimensional system without inversion symmetry. 
We establish a method to calculate the current-induced orbital magnetization in the in-plane direction by using real-space coordinates in the thickness direction.
By applying this theory to surfaces and interfaces of insulators, we find that an electric current along surfaces and interfaces induces an orbital magnetization perpendicular to the electric current.
\end{abstract}

\maketitle

\section{Introduction} 
In recent years, spintronics in which both electron spin and charge degrees of freedom are utilized, has made remarkable progress.
This remarkable progress is supported by discoveries of various ways of conversions between a charge current and a spin current.
Famous typical examples of spin-charge conversion phenomena are the spin Hall effect \cite{SMN,JSD,KMG,JSS} and the Edelstein effect \cite{VME,JII,KMG1,VSR}.
In the spin Hall effect, an electric current generates a transverse spin current in a material with spin-orbit interaction.
Conversely, a spin current can generate a transverse charge current by the inverse spin Hall effect.

In addition to the spin Hall effect, the Edelstein effect is also known as a spin-charge conversion phenomenon.
In the Edelstein effect, an charge current flowing through a material with spin-orbit interaction shifts the Fermi surface, and it produce a non-equilibrium spin polarization.
These effects are important for spintronics.

In ferromagnetic materials, in addition to the contribution of electron spin, there is another contribution to the magnetization of ferromagnetic materials, i.e. orbital magnetization.
In many ferromagnets, the spin magnetization is dominant and the orbital magnetization is small.
For example, in Fe, Ni, and Co, the orbital magnetization is 5\% to 10\% of the total magnetization \cite{RAR,AJG,DU}.
On the other hand, there are some examples in which orbital magnetization (OM) plays an important role, such as in weak ferromagnets where the spin and orbital magnetic moments are opposite to each other, as well as in magnetic nanowires, and in materials showing magnetoelectric coupling.
Therefore, understanding the orbital magnetism and mechanisms behind it can further enable us to develop spintronic and orbitronic devices, utilizing the orbital degrees of freedom.

Existence of the OM has been known for a long time, but until recently, there has been no method to calculate OM as a bulk quantity.
Then in 2005, the modern theory of OM was developed.
According to this theory, OM is expressed as an integral over the Brillouin zone (BZ) in terms of Bloch wave functions.
As such, it is determined by the band structure.
This formula can be derived from the semiclassical theory \cite{DX,DY,DX2}, the Wannier function approach \cite{RDC,TT,DC} and the perturbation theory \cite{JS}.
Furthermore, it has been reported that OM can be induced by an electric current through a chiral crystal \cite{TY2,TY}, but so far it applies only to a small class of systems.

In this paper, we construct a theory for the current-induced OM in systems without inversion symmetry.
In particular, we establish a method to calculate the current-induced OM in the in-plane direction, and we find that an electric current along surfaces and interfaces induces OM perpendicular to the electric current.
This approach can be widely applied to van der Waals atomic layered materials, surface and interfaces.
This paper is organized as follow.
In Sec.~I\hspace{-.1em}I, we describe our method for the calculation of the current-induced OM.
In Sec.~I\hspace{-.1em}I\hspace{-.1em}I, we apply this to a polar metal, and calculate the current-induced OM using a tight-binding model.
In Sec.~I\hspace{-.1em}V, we develop a new formula for the in-plane OM in the two-dimensional systems.
In Sec.~V, we show that the OM can be induced by an electric current on surfaces and interfaces of insulators.
In Sec.~V\hspace{-.1em}I, we summarize.

\section{Background: Current-induced orbital magnetization}
In this section, we derive a formula for the current-induced OM.
In a three-dimensional crystal, in the limit of zero temperature $T\to0$, the OM is calculated from the formula \cite{RDC,TT,DC,TY2,TY}
\begin{eqnarray}
[\bm{M}_{\mathrm{orb}}]_i&=&\frac{ie}{\hbar}\epsilon_{ijk}\sum_{n}\int_{\mathrm{BZ}}\frac{d^3k}{(2\pi)^3}f_{n\bm{k}}\sum_{m(\neq n)}\frac{2\epsilon_F-\epsilon_{n\bm{k}}-\epsilon_{m\bm{k}}}{(\epsilon_{n\bm{k}}-\epsilon_{m\bm{k}})^2} \nonumber \\
&\quad&\quad\times\bra{u_{n\bm{k}}}\frac{\partial \hat{H}_{\bm{k}}}{\partial k_j}\ket{u_{m\bm{k}}}\bra{u_{m\bm{k}}}\frac{\partial \hat{H}_{\bm{k}}}{\partial k_k}\ket{u_{n\bm{k}}}, \label{3.2.1}
\end{eqnarray}
where the integral is performed over the BZ, $n$ denotes the band index, $\ket{u_{n\bm{k}}}$ is the $n$-th eigenstate at the Bloch wavevector $\bm{k}$, $\epsilon_{n\bm{k}}$ is the corresponding eigenenergy, $f_{n\bm{k}}$ is the distribution function for the eigenenergy $\epsilon_{n\bm{k}}$, and $\epsilon_F$ is the Fermi energy.

If we focus on a nonmagnetic system with time reversal symmetry, Eq.~(\ref{3.2.1}) becomes zero in equilibrium because the integrand is an odd function of $\bm{k}$.
When a current is flowing in the system, the distribution function $f_{n\bm{k}}$ deviates from equilibrium, which may lead to OM.
Here we note that the integrand of Eq.~(\ref{3.2.1}) for OM is zero in systems with both time-reversal symmetry and inversion symmetry, and the current cannot induce the OM.
Therefore, since we have assumed time-reversal symmetry, we need to break inversion symmetry.
To induce the OM, we apply an electric field $\bm{E}$.
Within the Boltzmann approximation, the applied electric field along the $x$ axis $E_x$ changes $f_{n\bm{k}}$ into \cite{TY2,TY}
\begin{eqnarray}
f_{n\bm{k}}=f_{n\bm{k}}^0+eE_x\tau v_{n,x}\left.\frac{df}{d\epsilon}\right|_{\epsilon=\epsilon_{n\bm{k}}}, \label{3.2.2}
\end{eqnarray}
where $f_{n\bm{k}}^0=f(\epsilon_{n\bm{k}})$ is the Fermi distribution function in equilibrium, $\tau$ is the relaxation time assumed to be constant, and $v_{n,x}=(1/\hbar)\partial \epsilon_{n\bm{k}}/\partial k_x$ is the velocity in the $x$ direction.
Substituting $f_{n\bm{k}}$ into Eq.~(\ref{3.2.1}), we obtain the current-induced OM $\bm{M}_{\mathrm{orb}}$,
\begin{eqnarray}
[\bm{M}_{\mathrm{orb}}^{3D}]_i&=&\frac{ie^2E_x\tau}{\hbar}\epsilon_{ijk}\sum_{n}\int_{\mathrm{BZ}}\frac{d^3k}{(2\pi)^3}v_{n,x}\left.\frac{df}{d\epsilon}\right|_{\epsilon=\epsilon_{n\bm{k}}} \nonumber \\
&\quad&\times\sum_{m(\neq n)}\frac{2\epsilon_F-\epsilon_{n\bm{k}}-\epsilon_{m\bm{k}}}{(\epsilon_{n\bm{k}}-\epsilon_{m\bm{k}})^2} \nonumber \\
&\quad&\times \bra{u_{n\bm{k}}}\frac{\partial H_{\bm{k}}}{\partial k_j}\ket{u_{m\bm{k}}}\bra{u_{m\bm{k}}}\frac{\partial H_{\bm{k}}}{\partial k_k}\ket{u_{n\bm{k}}}. \label{9998}
\end{eqnarray}
Because of the time-reversal symmetry, the $f_{n\bm{k}}^0$ term in Eq.~(\ref{3.2.2}) does not contribute to $\bm{M}_{\mathrm{orb}}$.
In the $T\to0$ limit, $\partial f/\partial \epsilon$ has a sharp peak at $\epsilon_F$.
Therefore, for a band insulator, the induced OM $\bm{M}_{\mathrm{orb}}$ is zero.

\section{Polar metal}
In the previous work \cite{TY2,TY}, a current-induced OM in a helical crystal structure is calculated.
Meanwhile, from a symmetry viewpoint, this theory can also be applied to a broader class of systems without inversion symmetry.
In this section, we focus on polar metals, which have no inversion symmetry.
As a result, we find that a current-induced OM appears in a direction determined by the polar symmetry of the system. 

\subsection{What is a polar metal?}
Ferroelectrics are materials in which electric dipole moments are spontaneously aligned.
Electric polarization does not appear in metals because conduction electrons screen electric polarization.
Therefore, ferroelectrics are limited to insulators.
Many ferroelectrics are classified as displacement ferroelectrics.
This is non-polar in the high-temperature phase, and becomes polar in the low-temperature phase. 
Metals do not become ferroelectric, but they can undergo similar structural changes by lowering temperature.
In this sense, such a polar metal can be regarded as a ``ferroelectric" metal.

``Ferroelectric" metals were suggested by Anderson and Blount in their study on a structural phase transition of V$_3$Si\cite{AB}.
The transition is of the second order and it was described as martensitic from various experiments.
However, most of the already known martensitic phase transitions were of the first order and only characterized by a change in the shape of the unit cell.
Anderson and Blount tried to describe the phase transition of V$_3$Si using the Landau theory, with strain treated as the only parameter .
However, it proved impossible and was concluded that if the phase transition is of the second order, another unknown parameter except for strain is needed.
It was suggested that this unknown parameter is related to atomic displacement which breaks inversion symmetry globally.
From this study of the second-order phase transition of V$_3$Si, the idea of "ferroelectric'' metals was born.
This class of metals is known as polar metals, and there exist a number of materials \cite{AB,HS,BTB,MK}.

\subsection{Our model for a polar metal}
\begin{figure}[t]
  \includegraphics[width=8cm,clip]{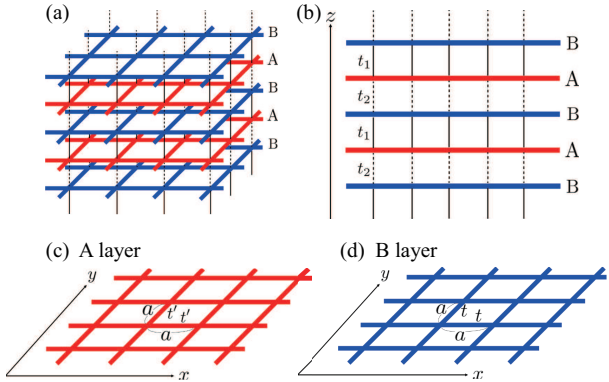}
  \caption{Our three-dimensional tight-binding model. (a) Lattice structure of our model of a polar metal. The B layers (blue) and the A layers (red) are stacked along the $z$-direction alternately. 
(b) Side view of our model. The two types of layers are connected by a hopping $t_1$ (solid lines) and $t_2$ (dashed lines) alternately. 
(c), (d) Two layers having a different hopping. The B layers (blue) have a hopping $t$ and the A layers (red) have a hopping $t'$.}
\label{zu1}
\end{figure}

In order to calculate a current-induced OM, we consider a three-dimensional tight-binding model of a polar metal as shown in Fig.~\ref{zu1}.
The lattice structure of this model is composed of an infinite stack of two types of square-lattice layers within the $xy$ plane: the A layers (red) having a nearest-neighbor hopping $t$ and the B layers (blue) having a nearest-neighbor hopping $t'$.
The two types of layers are stacked along the $z$-direction alternately and connected by nearest-neighbor interlayer hoppings $t_1$ (solid lines) and $t_2$ (dashed lines) alternately along the $z$ direction.
Here, $t$, $t'$, $t_1$ and $t_2$ are assumed to be real.
When $t\neq t'$ and $t_1\neq t_2$, this model does not have inversion symmetry.
Let $a$ and $c$ denote the lattice constants within the $xy$ plane and along the $z$ axis, respectively.
In the formula Eq.~(\ref{3.2.1}), we need to adopt a gauge \cite{RDC,TT}
\begin{eqnarray}
\begin{array}{l}
H_{\bm{k}+\bm{G}}=e^{-i\bm{G}\cdot\bm{r}}H_{\bm{k}}e^{i\bm{G}\cdot\bm{r}}, \\ 
\\
\ket{u_{n\bm{k}+\bm{G}}}=e^{-i\bm{G}\cdot\bm{r}}\ket{u_{n\bm{k}}}, \label{gauge}
\end{array}
\end{eqnarray}
where $\bm{G}$ is reciprocal lattice vector. 
We need to be careful because this gauge choice is different from the gauge $H_{\bm{k}+\bm{G}}=H_{\bm{k}}$ and  
$\ket{u_{n\bm{k}+\bm{G}}}=\ket{u_{n\bm{k}}}$, which is often adopted in various situations. 
Under this gauge, the Hamiltonian of this model is
\begin{eqnarray}
H_{\bm{k}}=\left(
\begin{array}{cc}
t\alpha & \beta^* \\
\beta & t'\alpha 
\end{array}
\right), \label{4.2.1}
\end{eqnarray}
where
\begin{eqnarray}
\alpha&=&2(\cos k_xa+\cos k_ya), \\
\beta&=&t_1e^{-ik_zc}+t_2e^{ik_zc} \label{abc}.
\end{eqnarray}
The unit cell consists of an atom in the A layers and an atom in the B layers, displaced along the $z$ direction.
The Hamiltonian contains the information of the position of atoms within unit cells. 
Namely, Eq.~(\ref{abc}) tells us that the A and B layers and displaced by $c$.

\begin{figure}[t]
  \includegraphics[width=9cm,clip]{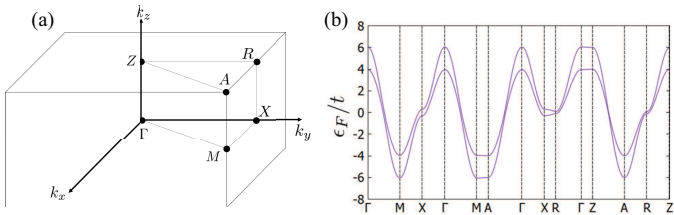}
  \caption{The Brillouin zone and band structure of our model. (a) The Brillouin zone.
(b) Band structure. Parameters are $t=1.0, t'=1.5t, t_1=0.1t$ and $t_2=0.2t$.
If $-6\leq\epsilon_F/t\leq6$, the system is a metal.}
\label{zu2}
\end{figure}

\subsection{Numerical result}
\begin{figure}[t]
  {\includegraphics[width=9cm,clip]{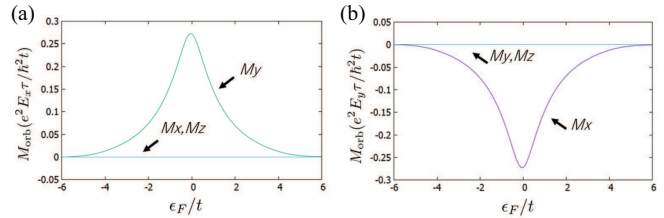}}
  \caption{Numerical result of current-induced OM in the model Eq.~({\ref{4.2.1}}) of a polar metal. Parameters are $t=1.0, t'=1.5t, t_1=0.1t$ and $t_2=0.2t$. The current-induced OM when the electric current is injected (a) in $x$ direction, and (b) in $y$ direction.}
\label{zu3}
\end{figure}

We compute the band structure of the model represented by the Hamiltonian Eq.~(\ref{4.2.1}), as shown in Fig.~\ref{zu2}.
Figure~\ref{zu2}(a) shows the Brillouin zone, and Fig.~\ref{zu2}(b) shows the band structure. 
If the Fermi energy $\epsilon_F$ lies in the energy band, this model is a metal and can carry electric current.
We calculate the current-induced OM by using Eq.~(\ref{3.2.1}) and Eq.~(\ref{3.2.2}).
Figure~\ref{zu3} shows the numerical result of the current-induced OM when a current passes along the $x$ direction and along the $y$ direction.
When the current flows along the $x$ direction (Fig.~\ref{zu3}(a)), the OM appears in the $+y$ direction, and likewise, when it flows along the $y$ direction (Fig.~\ref{zu3}(b)), the OM appears in the $-x$ direction.
The magnitudes of the current-induced OM in two cases are the same.

This numerical calculation reflects the symmetry of the model.
This model has $C_{4v}$ symmetry and a response tensor $\alpha_{ij}$ for the current-induced OM in a system with such symmetry is given by
\begin{eqnarray}
\bm{M}=\alpha\bm{J} \quad\Longrightarrow\quad \alpha=\left(
\begin{array}{ccc}
0 & -\alpha_{yx} & 0 \\ 
\alpha_{yx} & 0 & 0 \\
0 & 0 & 0
\end{array}
\right). \label{4.3.1}
\end{eqnarray} 
When an external field is applied along the $x$ direction, a response appears in the $+y$ direction, and likewise, when an external field is applied along the $y$ direction, a response with the same magnitude appears in the $-x$ direction.
Thus the result in Fig.~\ref{zu3} perfectly agrees with Eq.~(\ref{4.3.1}). 
Considering this result, in a general polar metal with the $z$ polarity direction, we can obtain the result that the OM appears in the direction perpendicular to the current.

\subsection{Interpretation of current-induced orbital magnetization}
Current-induced OM does not appear when the system has inversion symmetry.  
We investigate how the absence of inversion symmetry effects the current-induced OM by using the model of a polar metal.

We consider a case in which hopping parameters of electrons in the $xy$ plane are much larger than those along the $z$ direction;
\begin{eqnarray}
t_1,t_2\ll t,t'. \label{4.4.1}
\end{eqnarray}
Let us divide the Hamiltonian into two parts
\begin{eqnarray}
H_{\bm{k}}=H_0+V,
\end{eqnarray}
where
\begin{eqnarray}
H_0&=&\left(
\begin{array}{cc}
t\alpha & 0 \\
0 & t'\alpha
\end{array}
\right),\qquad V=\left(
\begin{array}{cc}
0 & \beta^* \\
\beta & 0
\end{array}
\right).
\end{eqnarray}
From our assumption Eq.~(\ref{4.4.1}), we have $|t\alpha|,|t'\alpha| \gg |\beta|$, and we regard the interlayer term $V$ as a perturbation term. 
We calculate current-induced OM by perturbation theory and we obtain
\begin{eqnarray}
M_y\propto t_1^2-t_2^2. \label{4.4.4}
\end{eqnarray}
Namely, the current-induced OM is proportional to a difference between two values of hoppings in $z$ direction.

\begin{figure}[t]
  {\includegraphics[width=9cm,clip]{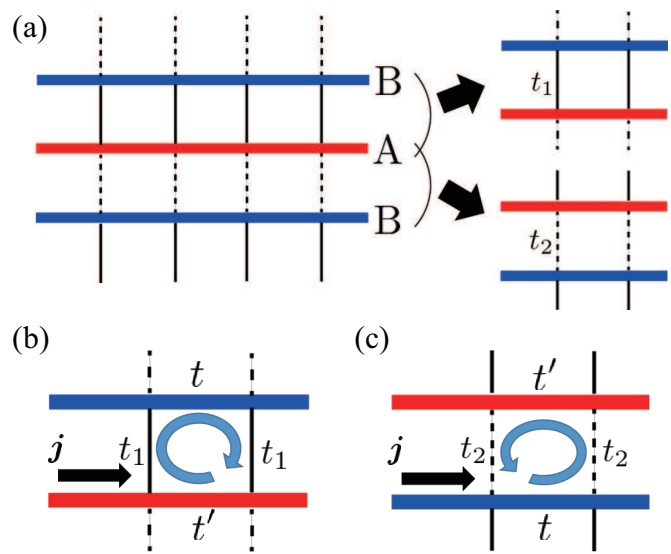}}
  \caption{Physical picture of the current-induced OM. (a) Two types of pairs of neighboring layers. We assume that the hopping ($t$) within the B layers (blue) is different from that ($t'$) within the A layers (red).
This model has two combinations of neighboring layers.
One is including $t_1$, and the other is including $t_2$.
(b)(c) Closed loops of an electric current. 
Due to this difference of the interlayer hoppings, we expect that the OM will appear.
The orientation of the current-induced OM within the two neighboring layers is reversed because the relative positions between $t$ and $t '$ is reversed.}
\label{zu4}
\end{figure}

We can give a physical interpretation for Eq.~(\ref{4.4.4}).
Electrons move along two different kinds of layers when the electric current flows in the $x$ direction.
Here, the in-plane velocities of electrons are different between the two layers since the in-plane hopping is different.
Due to this difference in velocities, when the current flows in the polar metal, electrons can be regarded to form a closed loop of an electric current in the crystal as shown in Fig.~\ref{zu4}, and we expect OM from this.
In this model, there are two types of closed loops, Fig.~\ref{zu4}(b) and Fig.~\ref{zu4}(c).
In a closed loop of Fig.~\ref {zu4}(b), the magnetization depends on the product of the hoppings on the bonds forming the closed loop of current;
\begin{eqnarray}
\bm{M}_{\mathrm{orb}}^{(\mathrm{b})}\propto tt't_1^2.
\end{eqnarray}
A closed loop of Fig.~\ref{zu4}(c) is treated similarly as in Fig.~\ref{zu4}(b), but the orientation of the OM is reversed because the relative positions between $t$ and $t '$ is reversed;
\begin{eqnarray}
\bm{M}_{\mathrm{orb}}^{(\mathrm{c})}\propto -tt't_2^2.
\end{eqnarray}
Thus, the total OM is proportional to $t_1^2-t_2^2$.
That is, the OM appears in a polar metal because closed loops of the current are formed in the crystal.
When $t_1\neq t_2$, inversion symmetry is broken and the OM appears in response to the current, because Fig.~\ref{zu4}(b) and Fig.~\ref{zu4}(c) are no longer equivalent.

\subsection{Polar metal: SnP}
\begin{figure}[t]
  {\includegraphics[width=9cm,clip]{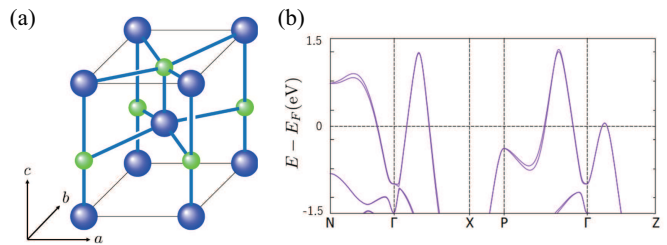}}
  \caption{Crystal structure and band structure of a polar metal SnP.
(a) Crystal structure of SnP in the polar phase. The blue spheres and the green spheres are Sn and P atoms, respectively.
The P atoms are displaced uniformly along the $c$-axis and SnP becomes a polar metal.
(b) Band structure of SnP in the polar phase, calculated from tight-binding model.
The spin-orbit interaction is small, and the band splitting is small.}
\label{zu5}
\end{figure}

In this subsection, we calculate the current-induced OM in a polar metal SnP known since half a century ago.
SnP crystalizes into the face-centered cubic lattice at room temperature at ambient pressure.
When the temperature drops below 250K at ambient pressure, the P atoms are displaced uniformly and SnP becomes a polar metal as shown in Fig.~\ref{zu5} \cite{MK}.

The space group of the polar metal SnP is $I4mm$, and we expect that the current-induced OM appears in direction perpendicular to the electric current, similar to the model in Sec.~I\hspace{-.1em}I\hspace{-.1em}IC.
{To study the current-induced OM in SnP, we first calculated the relativistic electronic structure of bulk SnP in the $I4mm$  phase within density functional theory (DFT) using Perdew-Burke-Ernzerhof exchange-correlation functional as implemented in WIEN2K program~\cite{wien2k}. For all atoms, the muffin-tin radius $R_{MT}$ was chosen such that its product with the maximum modulus of reciprocal vectors $K_{max}$ become $R_{MT} K_{max}=7.0$.
The Brillouin zone was sampled using a $12\times 12\times 12$ $k$-mesh. From the DFT Hamiltonian we then downfolded a  $14\times14$ tight-binding model using maximally localized Wannier functions~\cite{souza, mostofi}.
The corresponding basis set was made of atomic spin-orbitals: P($p_x,\uparrow$), P($p_y,\uparrow$), P($p_z,\uparrow$), P($s, \uparrow$), Sn($p_x,\uparrow$), Sn($p_y,\uparrow$), Sn($p_z,\uparrow$), P($p_x,\downarrow$), P($p_y,\downarrow$), P($p_z,\downarrow$), P($s,\downarrow$), Sn($p_x,\downarrow$), Sn($p_y,\downarrow$) and Sn($p_z,\downarrow$) coming from the $3p$ orbitals of P atoms and the $5p$ orbitals of the Sn atoms.
We estimate the magnitude of the current-induced OM in SnP.
By assuming that the electric field is $E_x=10^4[\mathrm{V}/\mathrm{m}]$ and the life time is $\tau=10^{-12}[\mathrm{s}]$, the magnitude of the current-induced interatomic OM is about $B=0.63[\mathrm{G}]$, which can be measured in experiments.

\section{two-dimensional systems}
In the previous section, we showed that a current-induced OM appears in three-dimensional (3D) polar metals.
We found that the essence of current-induced OM is formation of closed loops of the electric current.
Therefore, we expect that it will also appear in a two-dimensional (2D) system without inversion symmetry.
From the results in the previous section, it is expected that current-induced OM will appear in the in-plane direction as well as along the out-of-plane direction, depending on symmetry of the system.
Here, we encounter a theoretical problem in calculating the in-plane OM in 2D systems.
As is clear from the formula of the OM, Eq.~(\ref{3.2.1}), information on the Bloch wave function in the $j$ and $k$ directions is required when calculating the OM in the $i$ direction, where $i, j$ and $k$ represent mutually perpendicular directions.
Actually, there has never been a known method to calculate in-plane OM in 2D systems, since in the direction perpendicular to the plane of a 2D system the wave functions are not of the Bloch form.

In this section, we develop a method to calculate the OM in the in-plane direction in a 2D system without inversion symmetry.
To this end, we proceed as follows.
First, we consider a hypothetical stacking of the 2D system to form a 3D system, for which one can calculate the OM by Eq.~(\ref{3.2.1}), and then we interpret this result as that of the original 2D system.
In the original 2D systems, one had the problem that the Bloch wave function in the direction perpendicular to the plane of a 2D system cannot be defined, and we can solve this problem by going through the 3D system.
Through this procedure, we derive a new formula of the in-plane OM using real-space coordinates in the out-of plane direction.
As a result, the OM in the in-plane direction of the 2D system can be calculated theoretically.

\subsection{Orbital magnetization in the in-plane direction}
As seen in the previous section, current-induced OM is caused by closed loops of a current in the crystal.
This is similar to the classical phenomenon of a magnetic field generated by a closed loop of a current.
These two phenomena show that a magnetic field or magnetization appears in a direction perpendicular to the plane along which the current flows.
In other words, to calculate OM in a certain direction, information of the spatial distribution of the wave function in two directions orthogonal to that direction is required.
This can also be seen from the formula of the OM (Fig.~(\ref{zu1})).
When calculating the OM in the $i$ direction, information on the Bloch wave function in the $j$ and $k$ directions is required, where the $i, j$ and $k$ directions are mutually perpendicular.  

Therefore, we encounter a problem when calculating the OM in the in-plane direction of a 2D system using Eq.~(\ref{3.2.1}).
As an example, we consider the OM in the $y$ direction in a 2D system along the $xy$ plane.
In a 3D system both $k_x$ and $k_z$ are well-defined, but in a 2D system along the $xy$ plane, $k_z$ cannot be defined because the system is finite along the $z$ direction.
This is a problem when calculating the in-plane OM of a 2D system.

\subsection{Calculation of the current induced orbital magnetization for 2D systems}
Here we solve this problem by going through a 3D system.
We extend the 2D system to a 3D system by putting the 2D system periodically along the $z$ direction without any hopping between the 2D systems.
Then the OM can be calculated since $k_z$ is well-defined.
The result can be regarded as the OM of the original 2D problem in the in-plane direction.

We explain this procedure with an example of the current-induced OM for a 2D model shown in Fig.~\ref{zu6}.
This model consists of two layers A and B, both forming a square lattice.
The A layers (red) have a nearest-neighbor hopping $t$, the B layers (blue) have a nearest-neighbor hopping $t'$ and the interlayer hopping is $t_1$ along the $z$-axis.
The Hamiltonian $\tilde{H}_{\tilde{\bm{k}}}$ of this model is
\begin{eqnarray}
\tilde{H}_{\tilde{\bm{k}}}=\left(
\begin{array}{cc}
t\alpha & t_1 \\
t_1 & t'\alpha
\end{array}
\right).
\end{eqnarray}

\begin{figure}[t]
  {\includegraphics[width=8cm,clip]{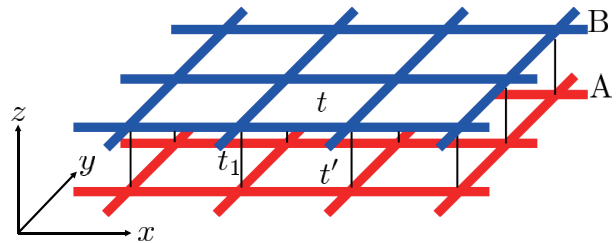}}
  \caption{Our 2D model. The red layer having a nearest-neighbor a hopping $t$ and the blue one having hopping $t'$ are connected by a hopping $t_1$. This bilayer model is finite along $z$ direction and infinite along $x, y$ directions.}
\label{zu6}
\end{figure}

We cannot calculate the current-induced OM in the $y$ direction by Eq.~(\ref{3.2.1}) since $k_z$ is not defined.
To resolve this problem, we stack the 2D systems periodically along the $z$ direction.
This system is exactly the same as the 3D model introduced in Sec.~I\hspace{-.1em}I\hspace{-.1em}IB, with $t_2=0$. 
Therefore, in the calculation we adopt the 3D Bloch Hamiltonian $H_{\bm{k}}$ in Eq.~(\ref{4.2.1}) with $t_2=0$ instead of $\tilde{H}_{\bm{k}}$, namely
\begin{eqnarray}
\tilde{H}_{\tilde{\bm{k}}}=\left(
\begin{array}{cc}
t\alpha & t_1 \\
t_1 & t'\alpha
\end{array}
\right) \,\,\Rightarrow\,\, H_{\bm{k}}=\left(
\begin{array}{cc}
t\alpha & t_1e^{ik_zc} \\
t_1e^{-ik_zc} & t'\alpha 
\end{array}
\right), \label{def}
\end{eqnarray}
where $\tilde{\bm{k}}=(k_x,k_y)$ and $\bm{k}=(k_x,k_y,k_z)$.
Next, we calculate a current-induced OM in the 3D system with $t_2=0$ in  Eq.~(\ref{9998}).
Finally, the OM as a 2D system is 
\begin{eqnarray}
\bm{M}_{\mathrm{orb}}^{2D}=2c\bm{M}_{\mathrm{orb}}^{3D},
\end{eqnarray}
because the period along the $z$ direction is $2c$.
Using this method, we can calculate current-induced OM $\bm{M}_{\mathrm{orb}}^{2D}$ in 2D systems.

We show the numerical result of the magnitude of the current-induced OM by changing the interlayer hopping  $t_2$ in Fig.~\ref{7}.
When $t_2/t_1=1$, the current-induced OM become zero since the 3D system has inversion symmetry.
When $t_2/t_1\to0$, the system becomes the original 2D system as we will discuss in the next subsection.
\begin{figure}[t]
  {\includegraphics[width=9cm]{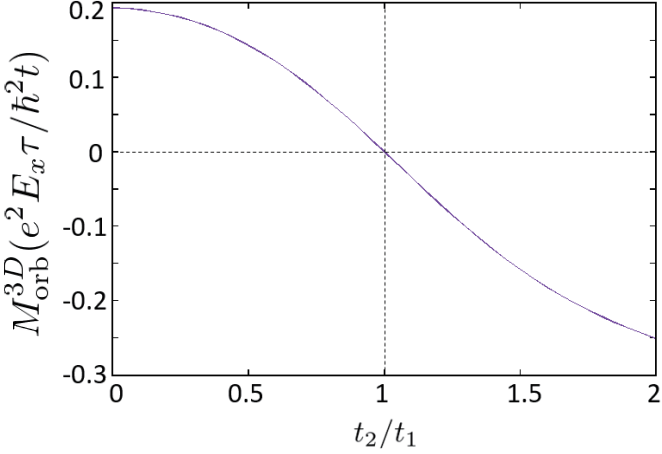}}
  \caption{Numerical calculation of the current-induced OM by changing $t_2$. The parameters are $t=1$, $t'=0.5$ and $\epsilon_F=1$. When $t_2/t_1=1$, a current-induced OM is zero because this model have inversion symmetry. When $t_2/t_1=0$, this model becomes a pure 2D system.}
\label{7}
\end{figure}

\subsection{Discussion}
We can calculate the in-plane OM of a 2D system by the method in the previous subsection.
This method is rather an indirect method since we should go through a 3D system.
Therefore, in this subsection we develop a method of calculating the in-plain OM of a 2D system without going through a 3D system.
In the present model, from Eq.~(\ref{def}), the 2D Hamiltonian $\tilde{H}_{\tilde{\bm{k}}}$ and the 3D Hamiltonian $H_{\bm{k}}$ are connected by unitary transformation;
\begin{eqnarray}
H_{\bm{k}}=U_{k_z}\tilde{H}_{\tilde{\bm{k}}}U_{k_z}^{\dagger},\quad U_{k_z}=\left(
\begin{array}{cc}
1 & 0 \\
0 & e^{-ik_zc}
\end{array}
\right). \label{5.4.1}
\end{eqnarray}
Therefore the eigenstates of the two Hamiltonians are connected by the unitary transformation $U_{k_z}$, which do not change their eigenvalues;
\begin{eqnarray}
\ket{u_{n\bm{k}}}&=&U_{k_z}\ket{\tilde{u}_{n\tilde{\bm{k}}}}, \label{5.4.2} \\
\epsilon_{n\bm{k}}&=&\tilde{\epsilon}_{n\tilde{\bm{k}}},
\end{eqnarray}
where $u_{n\bm{k}}, \tilde{u}_{n\tilde{\bm{k}}}$ and $\epsilon_{n\bm{k}}, \tilde{\epsilon}_{n\tilde{\bm{k}}}$ are the eigenstates and the eigenvalues of $H_{\bm{k}}$ and $\tilde{H}_{\tilde{\bm{k}}}$ respectively.
Therefore, in the limit $t_2\to0$ in the 3D model, the integrand of Eq.~(\ref{9998}) is transformed into
\begin{eqnarray}
&\,&\mathrm{Im}\left[\bra{u_{n\bm{k}}}\frac{\partial H_{\bm{k}}}{\partial k_z}\ket{u_{m\bm{k}}}\bra{u_{m\bm{k}}}\frac{\partial H_{\bm{k}}}{\partial k_x}\ket{u_{n\bm{k}}}\right] \nonumber \\
&\,&=\mathrm{Im}\left[i(\epsilon_{n\tilde{\bm{k}}}-\epsilon_{m\tilde{\bm{k}}})\bra{\tilde{u}_{n\tilde{\bm{k}}}}z\ket{\tilde{u}_{m\tilde{\bm{k}}}}\bra{\tilde{u}_{m\tilde{\bm{k}}}}\frac{\partial \tilde{H}_{\tilde{\bm{k}}}}{\partial k_x}\ket{\tilde{u}_{n\tilde{\bm{k}}}}\right], \nonumber \\ \label{5.4.3}
\end{eqnarray}
where $z=\mathrm{diag}(0,c)$ is the $z$ coordinate in this model.
Thus, while we start with the formula in three dimensions,
Eq.~(\ref{5.4.3}) can be used in two dimensions.
Because the term $\bra{u_{n\bm{k}}}\frac{\partial H_{\bm{k}}}{\partial k_z}\ket{u_{m\bm{k}}}$ becomes $i(\epsilon_{n\tilde{\bm{k}}}-\epsilon_{m\tilde{\bm{k}}})\bra{\tilde{u}_{n\tilde{\bm{k}}}}z\ket{\tilde{u}_{m\tilde{\bm{k}}}}$, we can get the OM from information on the real-space coordinates in the $z$ direction instead of information on $k_z$.

Therefore, the in-plane OM in general 2D systems along the $xy$ plane is
\begin{eqnarray}
&\,&[\bm{M}^{\mathrm{3D}}_{\mathrm{orb}}]_y=\frac{ie}{\hbar^2}\sum_{n}\int_{-\pi/2c}^{\pi/2c}\frac{dk_z}{2\pi}\int_{\mathrm{2DBZ}}\frac{d^2\tilde{\bm{k}}}{(2\pi)^2} \nonumber \\
&\,&\times f_{n\tilde{\bm{k}}}\sum_{m(\neq n)}\frac{2\epsilon_F-\epsilon_{n\tilde{\bm{k}}}-\epsilon_{m\tilde{\bm{k}}}}{(\epsilon_{n\tilde{\bm{k}}}-\epsilon_{m\tilde{\bm{k}}})^2} \nonumber \\
&\,&\times2i\mathrm{Im}\left[i(\epsilon_{n\tilde{\bm{k}}}-\epsilon_{m\tilde{\bm{k}}})\bra{\tilde{u}_{n\tilde{\bm{k}}}}z\ket{\tilde{u}_{m\tilde{\bm{k}}}}\bra{\tilde{u}_{m\tilde{\bm{k}}}}\frac{\partial \tilde{H}_{\tilde{\bm{k}}}}{\partial k_x}\ket{\tilde{u}_{n\tilde{\bm{k}}}}\right] \nonumber \\
&=&\frac{1}{2c}\frac{ie}{\hbar^2}\sum_{n}\int_{\mathrm{2DBZ}}\frac{d^2\tilde{\bm{k}}}{(2\pi)^2}f_{n\tilde{\bm{k}}}\sum_{m(\neq n)}\frac{2\epsilon_F-\epsilon_{n\tilde{\bm{k}}}-\epsilon_{m\tilde{\bm{k}}}}{(\epsilon_{n\tilde{\bm{k}}}-\epsilon_{m\tilde{\bm{k}}})^2} \nonumber \\
&\,&\times2i\mathrm{Im}\left[i(\epsilon_{n\tilde{\bm{k}}}-\epsilon_{m\tilde{\bm{k}}})\bra{\tilde{u}_{n\tilde{\bm{k}}}}z\ket{\tilde{u}_{m\tilde{\bm{k}}}}\bra{\tilde{u}_{m\tilde{\bm{k}}}}\frac{\partial \tilde{H}_{\tilde{\bm{k}}}}{\partial k_x}\ket{\tilde{u}_{n\tilde{\bm{k}}}}\right],\nonumber \\
\label{9999}
\end{eqnarray}
where $\int_{2DBZ}$ is an integral over the 2D BZ along the $k_x$-$k_y$ plane.
This shows that the term in the 3D system has been replaced as follows;
\begin{eqnarray}
\bra{u_{n\bm{k}}}\frac{\partial H_{\bm{k}}}{\partial k_z}\ket{u_{m\bm{k}}} \Longrightarrow i(\epsilon_{n\bm{k}}-\epsilon_{m\bm{k}})\bra{u_{n\bm{k}}}z\ket{u_{m\bm{k}}}.
\end{eqnarray}
This can be derived as
\begin{eqnarray}
\bra{u_{n\bm{k}}}\frac{\partial H_{\bm{k}}}{\partial k_z}\ket{u_{m\bm{k}}}&=&\bra{u_{n\bm{k}}}\hbar [\bm{v}_{\bm{k}}]_z\ket{u_{m\bm{k}}} \nonumber \\
&=&i\bra{u_{n\bm{k}}}\left[H_{\bm{k}},z\right]\ket{u_{m\bm{k}}} \nonumber \\
&=&i(\epsilon_{n\bm{k}}-\epsilon_{m\bm{k}})\bra{u_{n\bm{k}}}z\ket{u_{m\bm{k}}}.
\end{eqnarray}
This is consistent with the previous results.
Thus, the induced orbital magnetization per unit area of the 2D system is given by
\begin{eqnarray}
&\,&\left[\bm{M}_{\mathrm{orb}}^{2D}\right]_i=2c\left[\bm{M}_{\mathrm{orb}}^{3D}\right]_i \nonumber \\
&\,&\quad =\frac{ie}{\hbar^2}\epsilon_{jiz}\sum_{n}\int_{\mathrm{2DBZ}}\frac{d^2\tilde{\bm{k}}}{(2\pi)^2}f_{n\tilde{\bm{k}}}\sum_{m(\neq n)}\frac{2\epsilon_F-\epsilon_{n\tilde{\bm{k}}}-\epsilon_{m\tilde{\bm{k}}}}{(\epsilon_{n\tilde{\bm{k}}}-\epsilon_{m\tilde{\bm{k}}})^2} \nonumber \\
&\,&\quad \times2i\mathrm{Im}\left[i(\epsilon_{n\tilde{\bm{k}}}-\epsilon_{m\tilde{\bm{k}}})\bra{\tilde{u}_{n\tilde{\bm{k}}}}z\ket{\tilde{u}_{m\tilde{\bm{k}}}}\bra{\tilde{u}_{m\tilde{\bm{k}}}}\frac{\partial \tilde{H}_{\tilde{\bm{k}}}}{\partial k_j}\ket{\tilde{u}_{n\tilde{\bm{k}}}}\right]. \nonumber \\ \label{eq25}
\end{eqnarray}
While this formula is derived for the specific model in Fig.~\ref{zu6}, one can easily see that it applies to general 2D systems to calculate in-plane OM.  
We show the calculation result of the current-induced OM of our 2D model using Eq.~(\ref{9999}) in Fig.~\ref{zu8}.
For comparison, we also show the result using the 3D formula Eq.~(\ref{9998}).
This figure shows that the current-induced OM calculated by the two methods completely agree with each other.
\begin{figure}[t]
  {\includegraphics[width=9cm,clip]{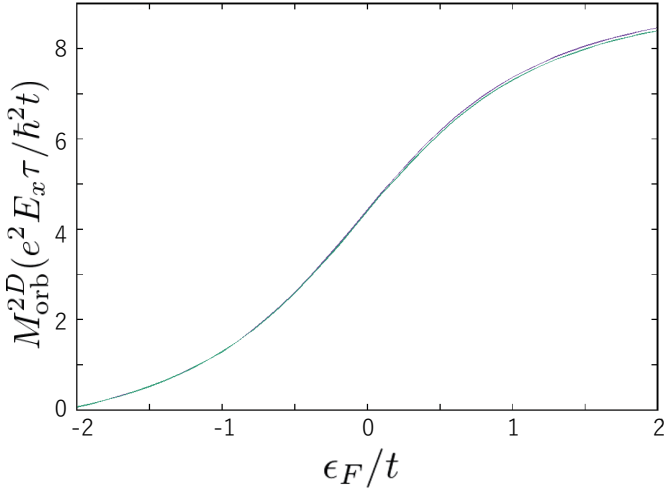}}
  \caption{Numerical results of the in-plane current-induced OM for the 2D model by the two methods. The parameters are $t=1,t'=0.5$ and $t_1=0.1$. The purple line is a numerical result going through 3D system, and the green line is a numerical result in the 2D system directly.
They completely agree with each other.}
\label{zu8}
\end{figure}

\subsection{General theory}
So far we have discussed the specific 2D model in Fig.~\ref{zu6}, and derived a formula for the current-induced orbital magnetization.
Here we show that Eq.~(\ref{eq25}) in fact applies to general 2D systems.
To show this, similarly to the previous subsection, we consider an arbitrary 2D system along $xy$ plane.
We then put the 2D system periodically along the $z$ direction without coupling them.
Then the original 2D Hamiltonian $\tilde{H}_{\tilde{\bm{k}}}$ and the resulting 3D Hamiltonian $H_{\bm{k}}$ are connected by the unitary matrix $U_{k_z}$,
\begin{eqnarray}
H_{\bm{k}}=U_{k_z}\tilde{H}_{\tilde{\bm{k}}}U_{k_z}^{\dagger},
\end{eqnarray}
where
\begin{eqnarray}
U_{k_z}=e^{-ik_zz}.
\end{eqnarray}
The eigenstates and eigenvalues of the two Hamiltonians satisfy the transformation,
\begin{eqnarray}
\ket{u_{n\bm{k}}}&=&U_{k_z}\ket{\tilde{u}_{n\tilde{\bm{k}}}}, \\
\epsilon_{n\bm{k}}&=&\tilde{\epsilon}_{n\tilde{\bm{k}}}.
\end{eqnarray}
These are consistent with our gauge choice Eq.~(\ref{gauge}).
Indeed, if we assume Eq.~(\ref{gauge}) for the original 2D system:
\begin{eqnarray}
\begin{array}{l}
\tilde{H}_{\tilde{\bm{k}}+\tilde{\bm{G}}}=e^{-i\tilde{\bm{G}}\cdot\tilde{\bm{r}}}\tilde{H}_{\tilde{\bm{k}}}e^{i\tilde{\bm{G}}\cdot\tilde{\bm{r}}}, \\
\ket{\tilde{u}_{n\tilde{k}+\tilde{G}}}=e^{-i\tilde{\bm{G}}\cdot\tilde{\bm{r}}}\ket{\tilde{u}_{n\tilde{\bm{k}}}}
\end{array}
\end{eqnarray}
where $\tilde{\bm{G}}=(G_x, G_y)$ and $\tilde{\bm{r}}=(x,y)$ are reciprocal lattice vector and real coordinate in $xy$ plane, 
then the corresponding 3D Bloch Hamiltonian and eigenstates satisfy the gauge condition in Eq.~(\ref{gauge}).  
Thus, the right-hand side of Eq.~(\ref{5.4.3}) is expressed as 
\begin{eqnarray}
&\,&\mathrm{Im}\left[\bra{u_{n\bm{k}}}\frac{\partial H_{\bm{k}}}{\partial k_z}\ket{u_{m\bm{k}}}\bra{u_{m\bm{k}}}\frac{\partial H_{\bm{k}}}{\partial k_x}\ket{u_{n\bm{k}}}\right] \nonumber \\
&\,&=\mathrm{Im}\left[(\epsilon_{m\tilde{\bm{k}}}-\epsilon_{n\tilde{\bm{k}}})\bra{\tilde{u}_{n\tilde{\bm{k}}}}U_{k_z}\frac{\partial U_{k_z}^{\dagger}}{\partial k_z}\ket{\tilde{u}_{m\tilde{\bm{k}}}}\bra{\tilde{u}_{m\tilde{\bm{k}}}}\frac{\partial \tilde{H}_{\tilde{\bm{k}}}}{\partial k_x}\ket{\tilde{u}_{n\tilde{\bm{k}}}}\right], \nonumber \\
&\,&=\mathrm{Im}\left[i(\epsilon_{m\tilde{\bm{k}}}-\epsilon_{n\tilde{\bm{k}}})\bra{\tilde{u}_{n\tilde{\bm{k}}}}z\ket{\tilde{u}_{m\tilde{\bm{k}}}}\bra{\tilde{u}_{m\tilde{\bm{k}}}}\frac{\partial \tilde{H}_{\tilde{\bm{k}}}}{\partial k_x}\ket{\tilde{u}_{n\tilde{\bm{k}}}}\right]. \nonumber \\ 
\end{eqnarray}
Hence the in-plane orbital magnetization is obtain as 
\begin{eqnarray}
&\,&[\bm{M}^{\mathrm{3D}}_{\mathrm{orb}}]_y=\frac{ie}{\hbar^2}\sum_{n}\int_{-\pi/d}^{\pi/d}\frac{dk_z}{2\pi}\int_{\mathrm{2DBZ}}\frac{d^2\tilde{\bm{k}}}{(2\pi)^2} \nonumber \\
&\,&\times f_{n\tilde{\bm{k}}}\sum_{m(\neq n)}\frac{2\epsilon_F-\epsilon_{n\tilde{\bm{k}}}-\epsilon_{m\tilde{\bm{k}}}}{(\epsilon_{n\tilde{\bm{k}}}-\epsilon_{m\tilde{\bm{k}}})^2} \nonumber \\
&\,&\times2i\mathrm{Im}\left[i(\epsilon_{n\tilde{\bm{k}}}-\epsilon_{m\tilde{\bm{k}}})\bra{\tilde{u}_{n\tilde{\bm{k}}}}z\ket{\tilde{u}_{m\tilde{\bm{k}}}}\bra{\tilde{u}_{m\tilde{\bm{k}}}}\frac{\partial \tilde{H}_{\tilde{\bm{k}}}}{\partial k_x}\ket{\tilde{u}_{n\tilde{\bm{k}}}}\right] \nonumber \\
&=&\frac{1}{d}\frac{ie}{\hbar^2}\sum_{n}\int_{\mathrm{2DBZ}}\frac{d^2\tilde{\bm{k}}}{(2\pi)^2}f_{n\tilde{\bm{k}}}\sum_{m(\neq n)}\frac{2\epsilon_F-\epsilon_{n\tilde{\bm{k}}}-\epsilon_{m\tilde{\bm{k}}}}{(\epsilon_{n\tilde{\bm{k}}}-\epsilon_{m\tilde{\bm{k}}})^2} \nonumber \\
&\,&\times2i\mathrm{Im}\left[i(\epsilon_{n\tilde{\bm{k}}}-\epsilon_{m\tilde{\bm{k}}})\bra{\tilde{u}_{n\tilde{\bm{k}}}}z\ket{\tilde{u}_{m\tilde{\bm{k}}}}\bra{\tilde{u}_{m\tilde{\bm{k}}}}\frac{\partial \tilde{H}_{\tilde{\bm{k}}}}{\partial k_x}\ket{\tilde{u}_{n\tilde{\bm{k}}}}\right],\nonumber \\
\end{eqnarray}
where $d$ is  the period along the $z$-direction.
Therefore, the in-plain orbital magnetization in the original 2D system, which is a magnetic dipole moment per unit area, is given by 
\begin{eqnarray}
&\,&\left[\bm{M}_{\mathrm{orb}}^{2D}\right]_i=l\left[\bm{M}_{\mathrm{orb}}^{3D}\right]_i \nonumber \\
&\,&\quad =\frac{ie}{\hbar^2}\epsilon_{jiz}\sum_{n}\int_{\mathrm{2DBZ}}\frac{d^2\tilde{\bm{k}}}{(2\pi)^2}f_{n\tilde{\bm{k}}}\sum_{m(\neq n)}\frac{2\epsilon_F-\epsilon_{n\tilde{\bm{k}}}-\epsilon_{m\tilde{\bm{k}}}}{(\epsilon_{n\tilde{\bm{k}}}-\epsilon_{m\tilde{\bm{k}}})^2} \nonumber \\
&\,&\quad \times2i\mathrm{Im}\left[i(\epsilon_{n\tilde{\bm{k}}}-\epsilon_{m\tilde{\bm{k}}})\bra{\tilde{u}_{n\tilde{\bm{k}}}}z\ket{\tilde{u}_{m\tilde{\bm{k}}}}\bra{\tilde{u}_{m\tilde{\bm{k}}}}\frac{\partial \tilde{H}_{\tilde{\bm{k}}}}{\partial k_j}\ket{\tilde{u}_{n\tilde{\bm{k}}}}\right], \nonumber \\
\end{eqnarray}
which is identical with Eq.~(\ref{eq25}).

Thus, we have established a formula for a direct calculation of the in-plane OM in a 2D system.
This method of calculating the in-plane OM in a 2D system can be applied to a wide range of systems.
It applies to the in-plane magnetization in equilibrium, as well as to that induced by a current.
Therefore, this formula applies to atomic-layer compounds and van der Waals heterostructures.

\section{Insulator surfaces and interfaces}
In the previous section, we found that OM is induced by a current in bulk 2D and 3D systems without inversion symmetry.
In this section, we focus on insulator surfaces and interfaces.
From symmetry arguments, it is expected that OM can be induced in such systems because inversion symmetry is broken.
First, we focus on bound states at an insulator surface, and consider that the Fermi energy lies in the surface band, making the surface metallic.
They exponentially decay away from the surface and can be characterized by a penetration depth.
We calculate the current-induced OM in the insulator surface using the formula in the previous section, and discuss its physical properties in this section.
If the penetration length is long, the region in which electrons can move is widened, and therefore current-induced OM is expected to increase.
However, if the penetration length is too long, the wave function has little influence from broken inversion symmetry, so that the current-induced OM is expected to decrease.

\subsection{Semiinfinite insulator with a surface}
We consider a tight-binding model of a semiinfinite insulator shown in Fig.~\ref{zu9}(a).
This model in Fig.~\ref{zu9}(a) expresses a semiinfinite system on a simple tetragonal lattice with a surface along the $xy$ plane having the $C_{4z}$ symmetry.
The lattice constant is $a$ along the $xy$ plane and $c$ along the $z$-axis.
Within each square-lattice layer within the $xy$ plane, the nearest-neighbor hopping is $-t$ and the on-site potential is $0$, except for the topmost layer where the nearest-neighbor hopping and the on-site potential are set to be $-t$ and $-\Delta$, respectively.
We assume $t'<t$.
The Hamiltonian of the tight-binding model is written as
\begin{eqnarray}
H_{\tilde{\bm{k}}}=\left(
\begin{array}{ccccc}
-\Delta-t'\alpha & -t & 0 & \cdots & \\
-t & -t\alpha & \ddots & \ddots & \\
0 & \ddots & \ddots & \ddots \\
\vdots & \ddots & \ddots & \ddots \\
\end{array}
\right).
\end{eqnarray}
The eigenstate $\ket{u_{\bm{k}}}$ of this Hamiltonian is expressed as
\begin{eqnarray}
\ket{u_{\bm{k}}}=\frac{1}{B}\left[\left(
\begin{array}{c}
1 \\
e^{-ik_zc} \\
e^{-2ik_zc} \\
\vdots 
\end{array}
\right)+A\left(
\begin{array}{c}
1 \\
e^{ik_zc} \\
e^{2ik_zc} \\
\vdots 
\end{array}
\right)\right], \label{66666}
\end{eqnarray}
where $B$ is a normalization factor, $A$ is given by
\begin{eqnarray}
A=-\frac{\Delta+(t'-t)\alpha-te^{ik_zc}}{\Delta+(t'-t)\alpha-te^{-ik_zc}},
\end{eqnarray}
and the eigenenergy is
\begin{eqnarray}
\epsilon_{\bm{k}}=-t\alpha-2t\cos(k_zc).
\end{eqnarray}
\begin{figure}[t]
  {\includegraphics[width=7cm]{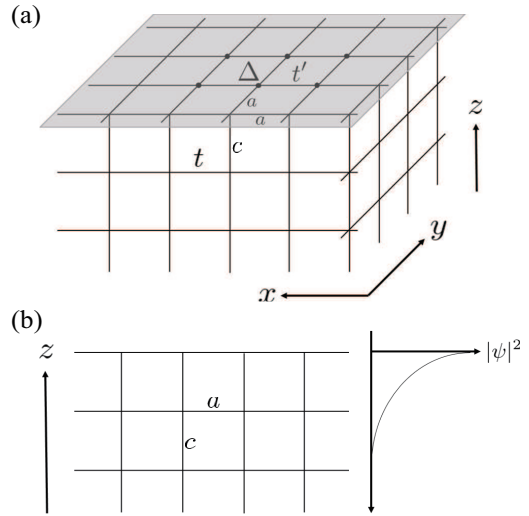}}
  \caption{(a) Our model of a semiinfinite insulator. The system is finite along $z$ direction and the lattice constant is $a$ along the $xy$ plane and $c$ along the $z$-axis.  
The hoppings are $-t$ and the on-site potential is $0$, but the hopping and the on-site potential of the surface are $-t'$ and $-\Delta$, respectively. (b) Schematic of the bound state. Electric density of the bound states decreases exponentially from the surface.}
\label{zu9}
\end{figure}

Here, we consider a bound state which is localized at the surface.
The wave function of the bound state decreases exponentially from the surface as shown in Fig.~\ref{zu9}(b).
In this situation, $k_z$ becomes a complex number; $k_z=-iK$ and $K>0$, and we impose $A=0$ because Eq. (\ref{66666}) should decay inside the insulator.
Therefore, the eigenvalue of the bound state becomes
\begin{eqnarray}
\begin{array}{l}
\epsilon_{\tilde{\bm{k}}}=-t\alpha-2t\cosh(Kc), \\
te^{Kc}=\Delta+(t'-t)\alpha(k_x,k_y).
\end{array}
\end{eqnarray}
Therefore existence of a bound state requires;
\begin{eqnarray}
\alpha<\frac{\Delta-t}{t-t'}. \label{6.1.10}
\end{eqnarray}
The bound states exist in the $k_x, k_y$ region satisfying Eq. (\ref{6.1.10}).
Another type of bound states is obtained by replacing $Kc$ with $Kc+\pi i$, namely
\begin{eqnarray}
\begin{array}{l}
\epsilon_{\tilde{\bm{k}}}=-t\alpha+2t\cosh(Kc), \\
-te^{Kc}=\Delta+(t'-t)\alpha(k_x,k_y),
\end{array}
\end{eqnarray}
and this type of bound states appear when 
\begin{eqnarray}
\alpha>\frac{\Delta+t}{t-t'}. 
\end{eqnarray}
We show the energy bands of this model along $k_x=k_y$ in Fig. \ref{zu6.1.3}.
The energy bands change by changing the on-site potential of the surface.
At $\Delta=0$, the bound states and the continuum band have an overlap in energy, and as the on-site potential increases, the bound states go down in energy.
At $\Delta=3$, the bound states and the continuum band are degenerate only at $\bm{k}=0$, and when $\Delta>3$, the bound states are completely separated from the continuum band.
\begin{figure}[t]
  {\includegraphics[width=8cm,clip]{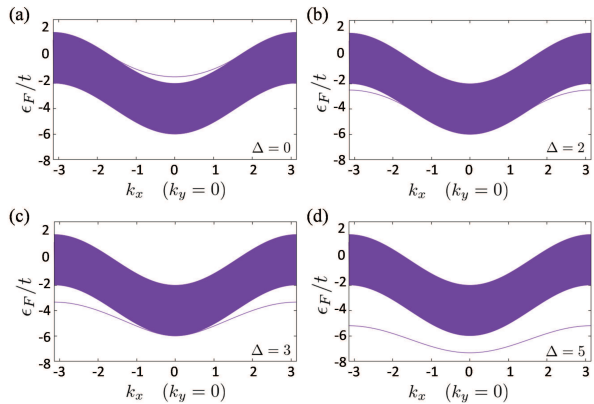}}
  \caption{The energy bands of this model for (a) $\Delta=0$, (b) $\Delta=2$ (c) $\Delta=3$ and (d) $\Delta=5$. The parameters are $t=1$ and $t'=0.5$. The bound states change when the on-site potential $\Delta$ on the insulator surface changes.
(a) At $\Delta=0$, the bound states are buried inside the continuum band.
(b), (c) At $\Delta=2$ and $\Delta=3$, as the on-site potential increases, the bound states go down in energy.
(d) At $\Delta=5$, the bound states are completely separated from the continuum band.}
\label{zu6.1.3}
\end{figure}

We calculate a current-induced OM when the Fermi energy lies within the band of the bound states but does not cross the continuum band, so that the current flows only along the surface.
The eigenenergy is written as
\begin{eqnarray}
\epsilon_{\tilde{\bm{k}}}=-\Delta-t'\alpha-\frac{t^2}{\Delta+(t'-t)\alpha}, 
\end{eqnarray}
and the eigenstate is given by
\begin{eqnarray}
u_{\tilde{\bm{k}}}=\frac{1}{B}\left(
\begin{array}{c}
1 \\
\displaystyle{\frac{t}{\Delta+(t'-t)\alpha}} \\
\displaystyle{\left(\frac{t}{\Delta+(t'-t)\alpha}\right)^2} \\
\vdots 
\end{array}
\right),
\end{eqnarray}
where the normalization factor $B$ is given by
\begin{eqnarray}
B=\left[1-\left(\frac{t}{\Delta+(t'-t)\alpha}\right)^2\right]^{-1/2}. 
\end{eqnarray}

\subsection{Numerical result}
We show the numerical result of the current-induced OM in Fig.~\ref{zu6.2.1} for $\Delta=5$, $t=1$ and $t'=0.5$, corresponding to the band structure shown in Fig.~\ref{zu6.1.3}(d).
Figure \ref{zu6.2.1} shows that the current-induced OM appears in an insulator surface.
Here, we assume that the Fermi energy crosses only the bound state, so that the current does not flow in the bulk but flow along the surface.
When $\epsilon_F/t$ is larger than the bottom of the surface band, the current-induced OM appears and its magnitude increases as $\epsilon_F$ increases.
This is because an area of the Fermi surface becomes larger.
The area of the Fermi surface corresponds to the number of conduction electrons in the surface band. 
If the area of the Fermi surface become larger, more electrons can move and the magnitude of the current-induced OM becomes larger.
\begin{figure}[t]
  {\includegraphics[width=8cm,clip]{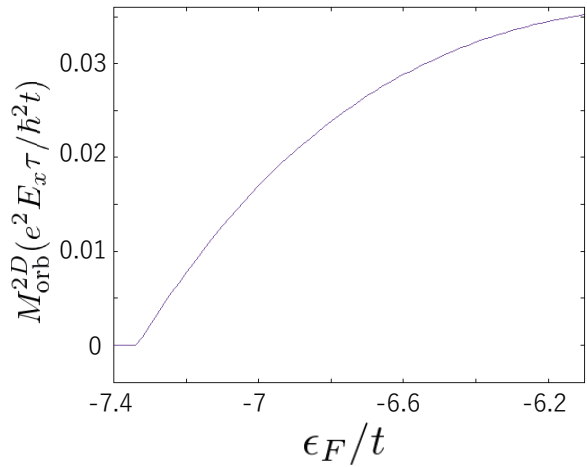}}
  \caption{Numerical result of the current-induced OM for an insulator surface.
The parameters are $\Delta=5, t=1$ and $t'=0.5$.
The orbital magnetization appears when $\epsilon_F/t \sim-7.3$ where the Fermi energy lies on the surface bound state in Fig.~\ref{zu6.1.3}(d) and the surface states are metallic.}
\label{zu6.2.1}
\end{figure}

\subsection{Discussion}
From the previous chapter, the formula of the current-induced OM in a system finite along $z$ direction is
\begin{eqnarray}
&\,&[\bm{M}^{\mathrm{2D}}_{\mathrm{orb}}]_i=\frac{ie^2E_z\tau}{\hbar}\sum_n\int_{\mathrm{2DBZ}}\frac{d^2\tilde{k}}{(2\pi)^2}\frac{\partial f_{n\tilde{\bm{k}}}}{\partial k_x}\epsilon_{ijz}\sum_{m(\neq n)}\nonumber \\
&\,&\frac{2\epsilon_F-\epsilon_{n\tilde{\bm{k}}}-\epsilon_{m\tilde{\bm{k}}}}{\epsilon_{n\tilde{\bm{k}}}-\epsilon_{m\tilde{\bm{k}}}} 2i\mathrm{Re}\left[\bra{u_{n\tilde{\bm{k}}}}z\ket{u_{m\tilde{\bm{k}}}}\bra{u_{m\tilde{\bm{k}}}}\frac{\partial H_{\tilde{\bm{k}}}}{\partial k_j}\ket{u_{n\tilde{\bm{k}}}}\right]. \nonumber \\ 
\end{eqnarray}
At zero temperature, because of the factor $\partial f_{n\tilde{\bm{k}}}/\partial k_x\propto \delta(\epsilon_{n\tilde{\bm{k}}}-\epsilon_F)$, we can put $\epsilon_{n\tilde{\bm{k}}}=\epsilon_F$ in the integrand. 
Moreover, using the completeness;
\begin{eqnarray}
\sum_{m(\neq n)}\ket{u_{m\bm{k}}}\bra{u_{m\bm{k}}}=1-\ket{u_{n\bm{k}}}\bra{u_{n\bm{k}}},
\end{eqnarray}
this formula becomes
\begin{eqnarray}
&\,&[\bm{M}^{\mathrm{2D}}_{\mathrm{orb}}]_i=\frac{ie^2E_z\tau}{\hbar}\int_{\mathrm{2DBZ}}\frac{d^2\tilde{k}}{(2\pi)^2}\frac{\partial f_{n\tilde{\bm{k}}}}{\partial k_x}\epsilon_{ijz} \nonumber \\
&\,&\times2i\mathrm{Re}\left[\bra{u_{n\tilde{\bm{k}}}}z\frac{\partial H_{\tilde{\bm{k}}}}{\partial k_j}\ket{u_{n\tilde{\bm{k}}}}-\bra{u_{n\tilde{\bm{k}}}}z\ket{u_{n\tilde{\bm{k}}}}\bra{u_{n\tilde{\bm{k}}}}\frac{\partial H_{\tilde{\bm{k}}}}{\partial k_j}\ket{u_{n\tilde{\bm{k}}}}\right]. \nonumber \\
\label{65432}
\end{eqnarray}

Next, by using this formula, we discuss how the magnitude of the current-induced OM depends on various parameters on surface states, by adopting some simple assumption.
In general, the wavefunctions of the surface states can be written as 
\begin{eqnarray}
u_{n\tilde{\bm{k}}}=\sqrt{1-e^{-2Kc}}\left(
\begin{array}{c}
1 \\
e^{-Kc} \\
e^{-2Kc} \\
\vdots 
\end{array}
\right), \label{6.3.1} 
\end{eqnarray}
where $K$ characterized the decay length of the surface states. 
Here, we adopt a convention where the position operator $z$ is a diagonal matrix, and each diagonal element expresses position information in a crystal;
\begin{eqnarray}
z=\mathrm{diag}(0, c, 2c, \cdots).
\end{eqnarray}
For simplicity, as is the case with our model, we assume that $\partial H_{\bm{k}}/\partial k_j$ is a real diagonal matrix.
We also assume that its diagonal component corresponding to the topmost layer is different from other diagonal components;
\begin{eqnarray}
\frac{\partial H_{\tilde{\bm{k}}}}{\partial k_j}=g_{\tilde{\bm{k}}}\mathrm{diag}(h_{\tilde{\bm{k}}}, 1, \cdots, 1),
\end{eqnarray}
with some functions $g_{\tilde{\bm{k}}}$ and $h_{\tilde{\bm{k}}}$. 
In addition, we assume that along the Fermi surface, the parameter $K$ characterizing the penetration depth is constant.
By inserting these assumptions into Eq.~(\ref{65432}), a current-induced OM becomes 
\begin{eqnarray}
[\bm{M}^{\mathrm{2D}}_{\mathrm{orb}}]_i\propto ce^{-2Kc}\int_{\text{Fermi surface}}d^2\tilde{k}g_{\tilde{\bm{k}}}(h_{\tilde{\bm{k}}}-1).
\end{eqnarray}
Therefore if $h_{\tilde{\bm{k}}}=1$, the resulting current-induced OM is always zero.
Namely, the modulation of the hopping at the layer near the surface is important within this model.
In other words, when $h_{\tilde{\bm{k}}}=1$, the velocity $v_j=\hbar^{-1}\frac{\partial H_{\tilde{\bm{k}}}}{\partial k_j}$ is proportional to an identity, and is diagonal in the basis of eigenstates $\ket{u_{n\tilde{\bm{k}}}}$, and Eq.~(\ref{65432}) vanishes. 
This means that the off-diagonal components of the velocity is essential for the current-induced OM in 2D systems.

The formulation of in-plane OM in insulator surfaces can also be applicable to interfaces between insulators.
At the interfaces the inversion symmetry is broken, making the interfaces to be polar.
Thus, when a current is flowing along the interfaces, an OM is induced perpendicularly to the current.
this theory can be applied to interfaces between insulators, such as SrTiO$_3$/LaAlO$_3$ interfaces.  

\section{conclusion}
In this paper, we discussed the current-induced OM in systems without inversion symmetry.
First, we focused on polar metals, which have no inversion symmetry in the bulk, and we showed that the OM is induced by an electric current in these systems.
Because of the crystal symmetry, the current-induced OM appears when the current is perpendicular to the polar direction, and the magnetization is perpendicular to both the current and the polar direction.
Moreover, using the perturbation theory, we also physically clarified how the current-induced OM appears in a polar metal.
From this result, the electric current forms closed loops in the crystal, and the OM is induced by them.
Our results can be generalized to any crystals without inversion symmetry. 
As an example, we calculated the current-induced OM in SnP, and showed that it might be experimentally measurable.

Second, we established a formula of the in-plain OM in a 2D system.
In a calculation of the OM in the in-plane direction, the known formula cannot be applied directly because the wavefunctions are not extended in the thickness direction.
We established a method to calculate the in-plane OM in a 2D system by virtually stacking the 2D system to form a 3D system.
This method can be applied to the current-induced OM in the in-plane direction.

Third, we discussed the current-induced OM on insulator surfaces  and interfaces by using our theory.
We showed that the current-induced OM can appear through surface states.
The phenomenon of the current-induced OM can be regarded as an orbital analog of the Edelstein effect, and therefore, it can be called orbital Edelstein effect.
In the conventional Edelstein effect, the spin-split bands by the spin-orbit coupling (SOC) \cite{VME,JII,KMG1,VSR}.
Thus, the SOC is needed for the conventional spin Edelstein effect.
In contrast, the orbital Edelstein effect does not require the SOC \cite{TY2,TY}.
We also note that the OM has two terms.
One is an intraatomic OM, due to the atomic orbitals with the angular momentum quantum number $l\geq1$, such as $p-$ and $d-$orbital.
The other is as interatomic OM, which is the main topic of our paper.  
In our paper, we only consider the latter contribution, and in real materials such as SnP, the other contribution due to the intraatomic OM should also be considered.

To summarize, current-induced OM is expected in a wide range of materials, and combination with \textit{ab initio} calculation can be a promising direction for future research.

\section*{acknowledgement}
This work is supported by JSPS KAKENHI Grant Number JP18H03678 and by the MEXT Elements Strategy Initiative to Form Core Research Center, Grant Number JPMXP0112101001.
MSB gratefully acknoledges support from CREST, JST  (Grant No. JPMJCR16F1).

\end{document}